\documentclass[aps,prd,preprintnumbers,superscriptaddress,nofootinbib,showpacs,twocolumn]{revtex4}%
\usepackage[dvipdfmx]{graphicx}
\usepackage{bm,latexsym,amsmath,amssymb,amsfonts,mathrsfs}
\usepackage{color}
\input{colordvi.tex}
\allowdisplaybreaks[1]
\usepackage[dvipdfmx]{hyperref}
\usepackage{url}
\hypersetup{
    colorlinks=true,
    citecolor=cyan,
}
\newcommand*{\D}{{\rm d}}
\newcommand*{\mpl}{M_{\rm Pl}}

\begin{document}

\title{Primordial tensor non-Gaussianities from general single-field inflation with non-Bunch-Davies initial states}

\author{Shingo~Akama}
\email[Email: ]{s.akama"at"rikkyo.ac.jp}
\affiliation{Department of Physics, Rikkyo University, Toshima, Tokyo 171-8501, Japan
}
\author{Shin'ichi Hirano}
\email[Email: ]{s.hirano"at"rikkyo.ac.jp}
\affiliation{Department of Physics, Rikkyo University, Toshima, Tokyo 171-8501, Japan
}
\author{Tsutomu~Kobayashi}
\email[Email: ]{tsutomu"at"rikkyo.ac.jp}
\affiliation{Department of Physics, Rikkyo University, Toshima, Tokyo 171-8501, Japan
}

\begin{abstract}
It has been found that the primordial non-Gaussianity of the curvature perturbation
in the case of non-Bunch-Davies initial states can be
enhanced compared with those in the case of the Bunch-Davies one
due to the interactions among the perturbations on subhorizon scales.
The purpose of the present paper is to
investigate whether tensor non-Gaussianities can also be enhanced or not
by the same mechanism.
We consider general gravity theory in the presence of an inflaton, and evaluate
the tensor auto-bispectrum and the cross-bispectrum involving
one tensor and two scalar modes
with the non-Bunch-Davies initial states for tensor modes.
The crucial difference from the case of the scalar auto-bispectrum is that
the tensor three-point function vanishes at the flattened momentum triangles.
We point out that the cross-bispectrum can potentially be enhanced
at non-trivial triangle shapes
due to the non-Bunch-Davies initial states.
\end{abstract}

\preprint{RUP-20-9}
\maketitle

\section{Introduction}
The validity of inflation~\cite{Guth:1980zm,Starobinsky:1980te,Sato:1980yn} is almost
beyond suspicion owing to both
the theoretical consistency that nicely
resolves the various problems in the standard Big Bang
cosmology and
the observational consistency with Planck and other data~\cite{Akrami:2018odb}.
Of particular interest is therefore to know which inflation model
is viable among a huge number of possibilities proposed so far.
One of the powerful tool for filtering inflationary models
is primordial
non-Gaussianity of the curvature perturbation, which turned
out to be not quite large according to
the observed CMB fluctuations~\cite{Akrami:2018odb} and
thus helped to rule out various models
predicting large non-Gaussianity. Similarly to this scalar non-Gaussianity,
the three-point functions involving tensor modes are expected to
have rich information about the physics of the early universe and
the interaction between gravity and the inflaton.


The inflationary perturbations have been
studied mainly by assuming the Bunch-Davies initial state~\cite{Bunch:1978yq}.
However, in principle,
the initial state is not necessarily given by the Bunch-Davies one,
and the validity of the assumptions on the initial state must be tested
against observations in the end.
Deviations from the Bunch-Davies state mean that the initial state is
excited, i.e., there exist particles initially.
In this case, the particles present initially can interact with each other
at early times, leading possibly to the generation of
non-Gaussianities on subhorizon scales.
Therefore, assuming non-Bunch-Davies initial states would
result in novel non-Gaussian signatures compared to
the standard case of the Bunch-Davies initial state.

The nature of primordial perturbations from non-Bunch-Davies initial states
has been explored so far in the literature~\cite{Chen:2006nt,Holman:2007na,Xue:2008mk,Meerburg:2009ys,Meerburg:2009fi,Chen:2010xka,Meerburg:2010ks,Chen:2010bka,Meerburg:2010rp,Agullo:2010ws,Ashoorioon:2010xg,Ganc:2011dy,Ganc:2012ae,Agarwal:2012mq,Gong:2013yvl,Flauger:2013hra,Aravind:2013lra,Ashoorioon:2013eia,Brahma:2013rua,Bahrami:2013isa,Kundu:2013gha,Emami:2014tpa,Zeynizadeh:2015eia,Meerburg:2015yka,Ashoorioon:2016lrg,Shukla:2016bnu,Ashoorioon:2018sqb}.
In particular, it has been found that the non-Gaussianity of
the curvature perturbation at the squeezed and flattened configurations
can be enhanced compared with those in the case of the Bunch-Davies state~\cite{Holman:2007na,Xue:2008mk,Meerburg:2009ys,Meerburg:2009fi,Chen:2010bka,Ashoorioon:2010xg,Ganc:2011dy,Gong:2013yvl,Bahrami:2013isa,Shukla:2016bnu}.
It is therefore natural to ask whether or not the
non-Gaussianities associated with the tensor modes can be enhanced as well.
There have already been several studies regarding
tensor non-Gaussianities from non-Bunch-Davies initial states~\cite{Kundu:2013gha,Ashoorioon:2018sqb}.
To address this question in more detail, in this paper
 we investigate the auto-bispectra of tensor
 modes and the cross-bispectra
 involving one tensor and two scalar modes in more general gravity
 theory than in the previous literature.

One naively expects that higher-derivative interactions have more impacts on
non-Gaussianities due to non-Bunch-Davies initial states.
Generalizing the underlying gravity theory yields such higher-derivative
interactions.
As a framework including higher derivative interactions,
we use an effective description of scalar-tensor gravity,
writing down the operators composed of the geometrical quantities such as
extrinsic and intrinsic curvature tensors~\cite{Gleyzes:2014dya,Gao:2014soa}.
Based on this effective description, in the present paper
we will estimate the size of tensor non-Gaussianities
from non-Bunch-Davies initial states
in general single-field inflation models.

This paper is outlined as follows. In the next section,
we consider general quadratic and cubic actions for tensor modes
and introduce non-Bunch-Davies initial states from a Bogoliubov transform
of the usual Bunch-Davies modes.
In Sec.~III, we calculate the auto-bispectrum of the tensor modes,
and investigate whether the enhanced non-Gaussian amplitudes can be obtained or not.
We then study in Sec.~IV the cross-bispectrum involving one tensor and two scalar modes,
and discuss how it can be enhanced compared with the case of the Bunch-Davies initial state.
A summary of the present paper is given in Sec.~V.

\section{Tensor modes}

\subsection{General quadratic and cubic interactions}

In the present paper, we investigate the properties of the tensor modes with non-Bunch-Davies initial states in order to see whether the tensor non-Gaussianities
could be enhanced or not.
Although we also study the cross-bispectrum with
the scalar modes briefly,
here we only summarize the quadratic and cubic interactions of
the tensor modes.

To derive the generic action for the tensor modes during inflation,
it is convenient to
employ the ADM decomposition with uniform inflaton hypersurfaces as
constant time hypersurfaces and
write down the possible
operators composed of the extrinsic curvature tensor $K_{ij}$ and
the intrinsic curvature tensor $R^{(3)}_{ij}$ of the constant time
hypersurfaces. First, the operators having the dimension
of mass squared are
\begin{align}
\mathcal{L}_{\rm GR}\supset K_{ij}^2,\ K^2,\ R^{(3)}, \label{mass2}
\end{align}
where $K$ is the
trace of $K_{ij}$. All these terms are present in general relativity.
Then, one can consider the leading-order corrections to Eq.~(\ref{mass2}):
\begin{align}
\mathcal{L}_{\rm cor}&\supset K_{ij}^3,\ K K_{ij}^2,\ K^3,\ K^{ij}R^{(3)}_{ij},\ KR^{(3)}. \label{mass3}
\end{align}
One may anticipate that
these corrections play a crucial role in
the generation of non-Gaussianities.
Therefore, in the present study, we consider the Lagrangian up to
this order, and
evaluate the contributions on the non-Gaussian amplitudes
from these correction terms.

More specifically, we consider the following wide class of the ADM action,
\begin{align}
S=\int\D t \D^3x \sqrt{\gamma}N\mathcal{L}, \label{massS}
\end{align}
where $\gamma$ is the determinant of the spacial metric $\gamma_{ij}$,
$N$ is the lapse function, and
\begin{align}
\mathcal{L}&=M_0^4(t,N)+M_1^3(t,N)K+M_2^2(t,N)\left(K^2-K_{ij}^2\right)\notag\\
&\quad+M_3^2(t,N)R^{(3)}+M_4(t,N)\left(K^3-3KK_{ij}^2+2K_{ij}^3\right)\notag\\
&\quad+M_5(t,N)\biggl(K^{ij}R^{(3)}_{ij}-\frac{1}{2}KR^{(3)}\biggr), \label{massL}
\end{align}
with $M_i(t,N)$ being a function
having the dimension of mass.
Here we have included the lower-order terms $M_0^4$ and $M_1^3K$,
though they do not contribute to the action for the tensor modes.
Equation~\eqref{massL} is nothing but the so-called GLPV Lagrangian~\cite{Gleyzes:2014dya},
and it includes the Horndeski theory as a subclass.
By introducing a St\"{u}ckelberg field
$\phi$, one can restore the full 4D covariance.

The transverse and traceless tensor perturbations $h_{ij}$ on top of
a spatially flat Friedmann-Lema\^{i}tre-Robertson-Walker
background are defined by
\begin{align}
\D s^2=-N^2(t)\D t^2+\gamma_{ij}\D x^i\D x^j,
\quad \gamma_{ij}=a^2(t)\left(e^h\right)_{ij},
\end{align}
where
\begin{align}
\left(e^h\right)_{ij}&:=\delta_{ij}+h_{ij}+
\frac{1}{2}h_{ik}h^k_j+\frac{1}{6}h_{ik}h^k_lh^l_j+\cdots.
\end{align}
At the level of the background, we may always
reparameterize the time coordinate so that we hereafter take $N=1$
and write $M_i(t,N(t))=M_i(t)$.
Since $\sqrt{\gamma}$ and
the trace part $K$ do not involve $h_{ij}$,
the terms such as
$M_0^4,\ M_1^3K$, and $M_2^2K^2$
in the Lagrangian do not contribute to the dynamics of the tensor perturbations.

Substituting the above metric into Eq.~(\ref{massL}),
the action for the tensor perturbations up to cubic order in $h_{ij}$
can be obtained as~\cite{Kobayashi:2011nu,Gao:2011vs}
\begin{align}
S_h=S^{(2)}_h+S^{(3)}_h,
\end{align}
where
\begin{align}
S^{(2)}_h&=\int{\D}t{\D}^3xa^3\frac{M_T^2}{c_h^2}\left[{\dot{h}_{ij}}^2-\frac{c_h^2}{a^2}(\partial_kh_{ij})^2\right], \label{2ndc}\\
S^{(3)}_h&=\int\D t\D^3xa^3\biggl[\frac{M_T^2}{4a^2}\left(h_{ik}h_{jl}-\frac{1}{2}h_{ij}h_{kl}\right)\partial_k\partial_lh_{ij}\notag\\
&\quad\quad\quad\quad\quad\quad\ +\frac{M_4}{4}\dot{h}_{ij}\dot{h}_{jk}\dot{h}_{ki}
\biggr] \label{3rdS}\\
&=:-\int\D tH_{\rm int},\notag
\end{align}
with
\begin{align}
M_T^2&:=2M_3^2+\dot M_5,\\
c_h^2&:=-2(M_2^2+3HM_4)/M_T^2.
\end{align}
Here, a dot stands for differentiation with respect to $t$ and $H:=\dot a/a$.
The interaction Hamiltonian $H_{\rm int}$
is introduced for later convenience.
We assume that $M_T\sim M_{\rm Pl}$.
The terms in the first line in Eq.~(\ref{3rdS})
are present in general relativity, while
the one in the second line
is a new operator introduced as a result of the extension of general
relativity with $M_4(t)\neq0$.
For example, this operator is obtained from the so-called $G_5$ term
in the Horndeski theory~\cite{Gao:2011vs}.
One might think that the third line in Eq.~(\ref{massL}) could also lead to
a new cubic operator, but it turns out that this can
be integrated by parts to yield the same terms as
in the first line in Eq.~(\ref{3rdS}).

\subsection{Non-Bunch-Davies initial states}

We now move to the Fourier domain,
\begin{align}
h_{ij}(t,{\bf x})=\int\frac{\D^3k}{(2\pi)^3}\tilde{h}_{ij}(t,{\bf k})e^{i{\bf k}\cdot{\bf x}}.
\end{align}
In the standard setup,
one expands the quantized tensor modes as
\begin{align}
\tilde{h}_{ij}(t,{\bf k})&=\sum_{s}\left[u_k(t)e^{(s)}_{ij}({\bf k})a^{(s)}_{\bf k}+u^*_k(t)e^{(s)*}_{ij}(-{\bf k})a^{(s)\dagger}_{-{\bf k}}\right], \label{Four-BD}
\end{align}
where $u_k(t)$ is the Bunch-Davies mode function.
The creation and annihilation operators satisfy
\begin{align}
\left[a^{(s)}_{\bf k},a^{(s')\dagger}_{\bf k'}\right]&=(2\pi)^3\delta_{ss'}\delta\left({\bf k}-{\bf k'}\right), \label{cr-an1}\\
{\rm{others}}&=0, \label{cr-an2}
\end{align}
and the subscript $s$ denoting two helicity modes takes $s=\pm$.
The polarization tensor, $e_{ij}$, satisfies the transverse and traceless conditions, $\delta_{ij}e^{(s)}_{ij}({\bf k})=0=k^ie^{(s)}_{ij}(\bf k)$.
This also satisfies $e^{(s)}_{ij}({\bf k})e^{(s')*}_{ij}({\bf k})=\delta_{ss'}$
and $e^{(s)*}_{ij}({\bf k})=e^{(-s)}_{ij}({\bf k})=e^{(s)}_{ij}(-{\bf k})$.

The equation of motion for the mode function $u_k$ is
derived from Eq.~\eqref{2ndc} as
\begin{align}
\frac{\D}{\D t}\left(\frac{a^3M_T^2}{c_h^2}\dot u_k\right)+aM_T^2k^2u_k=0. \label{eq-mode}
\end{align}
We solve Eq.~\eqref{eq-mode}
under the assumption that $M_T^2,\ c_h^2={\rm const}$
in the de Sitter background, $H={\rm const}$.
Then, the Bunch-Davies mode function is obtained as
\begin{align}
u_k=\frac{\sqrt \pi}{a}\frac{c_h}{M_T}\sqrt{-\eta}H^{(1)}_{3/2}(-c_hk\eta), \label{mode}
\end{align}
where $H^{(1)}_{3/2}$ is the Hankel function of the first kind of order $3/2$.
We write the state annihilated by ${\hat a}^{(s)}_{\bf k}$
as $|0_a\rangle$: ${\hat a}^{(s)}_{\bf k}|0_a\rangle=0$.

In this paper, we instead expand $\tilde h_{ij}$ as
\begin{align}
\tilde{h}_{ij}=\sum_{s}\left[\psi^{(s)}_ke^{(s)}_{ij}({\bf k})b^{(s)}_{\bf k}+\psi^{(s)*}_k{e^{(s)}_{ij}}^*(-{\bf k})b^{(s)\dagger}_{-{\bf k}}\right], \label{nBD-h}
\end{align}
where $\psi_k^{(s)}$ is a Bogoliubov transform of
the Bunch-Davies modes,
\begin{align}
\psi^{(s)}_k=\alpha^{(s)}_ku_k+\beta^{(s)}_ku^*_k. \label{nBD-mode}
\end{align}
The Bogoliubov coefficients are normalized as
$|\alpha^{(s)}_k|^2-|\beta^{(s)}_k|^2=1$
and the creation and annihilation operators satisfy
\begin{align}
a^{(s)}_{\bf k}&=\alpha^{(s)}_kb^{(s)}_{{\bf k}}+{\beta^{(s)}_k}^*b^{(s)\dagger}_{-{\bf k}},\\
b^{(s)}_{\bf k}&=\alpha^{(s)*}_{k}a^{(s)}_{\bf k}-\beta^{(s)*}_ka^{(s)\dagger}_{-{\bf k}}.
\end{align}
We write the state annihilated by $b^{(s)}_{\bf k}$ as
$|0_b\rangle$:
\begin{align}
b^{(s)}_{\bf k}|0_b\rangle=0.
\end{align}
Nonvanishing $\beta^{(s)}_k$ coefficients
indicate that the tensor modes get excited from
the Bunch-Davies vacuum, $a^{(s)}_\mathbf{k}|0_b\rangle\neq0$.

Let us assume
that the deviations from the Bunch-Davies initial
states are characterized by some small, real parameters as
\begin{align}
\beta^{(s)}_k&=\delta^{(s)}_1(k)+i\delta^{(s)}_2(k), \label{beta}\\
\alpha^{(s)}_k&=1+i\delta^{(s)}_3(k), \label{alpha}
\end{align}
where $\delta^{(s)}_1\sim\delta^{(s)}_2\sim\delta^{(s)}_3 \ll1$.
This is a reasonable assumption because
the magnitude of $\beta^{(s)}_k$ has an upper bound
in order for the inflationary background not to be spoiled by the excited tensor
modes, which is typically given by $|\beta^{(s)}_k|\lesssim 10^{-6}$
as argued in Appendix~\ref{app}.
The assumption on the form of $\alpha_k^{(s)}$
[Eq.~\eqref{alpha}] follows from $|\alpha^{(s)}_k|^2-|\beta^{(s)}_k|^2=1$.

\subsection{Primordial power spectrum}

The two-point correlation function is defined by
\begin{align}
\langle0_b|\tilde{h}_{ij}({\bf k})\tilde{h}_{kl}({\bf k'})|0_b\rangle
=(2\pi)^3\delta^{(3)}({\bf k}+{\bf k}')\mathcal{P}_{ij,kl},
\end{align}
where
\begin{align}
\mathcal{P}_{ij,kl}:=\sum_{s,s'}\biggl[\psi^{(s)}_k\psi^{(s')*}_ke^{(s)}_{ij}({\bf k})e^{(s')*}_{kl}({\bf k})\biggr].
\end{align}
Using Eqs.~(\ref{mode})--(\ref{nBD-mode}),
we obtain the power spectrum $\mathcal{P}_h$ as
\begin{align}
\mathcal{P}_h:=\frac{k^3}{2\pi^2}\mathcal{P}_{ij,ij}=\frac{1}{\pi^2}\frac{H^2}{M_T^2c_h}
\sum_{s}
\left|\alpha_k^{(s)}-\beta_k^{(s)}\right|^2, \label{power}
\end{align}
evaluated at the time of horizon crossing, $c_hk=aH$.
Its tilt is then derived as
\begin{align}
n_t&:=\frac{{\rm d}\ln \mathcal{P}_h}{{\rm d}\ln k}\notag\\
&\ \simeq-2\epsilon-s_h-2m_T+\frac{\rm d}{{\rm d}\ln k}\sum_{s}\left|\alpha^{(s)}_k-\beta^{(s)}_k\right|^2, \label{tilt-sr}
\end{align}
where
\begin{align}
\epsilon:=-\frac{\dot H}{H^2},
\quad  s_h:=\frac{\dot c_h}{Hc_h},\quad  m_h:=\frac{\dot{M}_T}{HM_T}, \label{slow}
\end{align}
are assumed to be small.
To leading order in $\beta_k^{(s)}$ we have
$|\alpha^{(s)}_k-\beta^{(s)}_k|^2\simeq 1-2{\rm Re}[\beta_k^{(s)}]$,
and so
\begin{align}
n_t\simeq -2\epsilon-s_h-2m_T-2\sum_s
\frac{\D\, {\rm Re}[\beta_k^{(s)}] }{\D\ln k}.\label{tilt-sr2}
\end{align}
This is a rather straightforward generalization of
previous results,
simultaneously taking into account the different effects on the spectral tilt:
the time variation of the inflationary Hubble parameter,
the speed of gravitational waves, and the effective Planck mass,
as well as the $k$-dependence of the Bogoliubov coefficients.
Note that in principle the sign of each term in Eq.~\eqref{tilt-sr2} is
not constrained. In particular, a blue tensor spectrum can be obtained
as a consequence of a time-dependent speed of gravitational
waves~\cite{Giovannini:2015kfa,Cai:2015yza,Cai:2016ldn,Giovannini:2018zbf,Giovannini:2018nkt}
and/or $k$-dependent $\beta_k^{(s)}$~\cite{Ashoorioon:2014nta}
even if the null energy condition is preserved, $\epsilon > 0$.

\section{Auto-bispectrum}

Let us now calculate the tensor three-point correlation functions with
non-Bunch-Davies initial states. Since the cubic interaction~\eqref{3rdS}
is composed of the two different contributions, i.e., the one
present in general relativity and the new one beyond general relativity,
we write the bispectrum as
$\mathcal{B}^{s_1s_2s_3}_{({\rm GR})}+\mathcal{B}^{s_1s_2s_3}_{({\rm new})}$,
where $\mathcal{B}^{s_1s_2s_3}_{\rm (GR)}$ and $\mathcal{B}^{s_1s_2s_3}_{\rm (new)}$
are originated from the former and the latter, respectively.
The tensor bispectrum (evaluated at $t=t_f$) is defined by
\begin{align}
&\langle0_b|\xi^{s_1}(t_f,{\bf k}_1)
\xi^{s_2}(t_f,{\bf k}_2)\xi^{s_3}(t_f,{\bf k}_3)|0_b\rangle\notag\\
&\ =(2\pi)^3\delta({\bf k}_1+{\bf k}_2+{\bf k}_3)\left(\mathcal{B}^{s_1s_2s_3}_{({\rm GR})}+\mathcal{B}^{s_1s_2s_3}_{({\rm new})}\right), \label{3-pt}
\end{align}
with $\xi^{s}(t,{\bf k}):=\tilde{h}_{ij}(t,{\bf k})e^{(s)*}_{ij}({\bf k})$, and
the three-point correlation function can be calculated using the in-in formalism
as
\begin{align}
&\langle0_b|\xi^{s_1}(t_f,{\bf k}_1)\xi^{s_2}(t_f,{\bf k}_2)\xi^{s_3}(t_f,{\bf k}_3)|0_b\rangle\notag\\
& =-i\int^{t_f}_{t_i}\D t'\langle[\xi^{s_1}(t_f,{\bf k}_1)\xi^{s_2}(t_f,{\bf k}_2)
\xi^{s_3}(t_f,{\bf k}_3),H_{{\rm int}}(t')]\rangle. \label{in-in}
\end{align}
Here, $t_i$ is some time when the perturbation modes
are deep inside the horizon, and $t_f$
is the time at the end of inflation.
In terms of the conformal time defined by $\D\eta:=\D t/a$,
we take $\eta_f=0$. As for the initial time,
we do not simply take $\eta_0\to-\infty$, but we keep it finite,
$\eta_i=\eta_0\,(<0)$, where $\eta_0$ is associated with
the cutoff scale $M_*$ as
$M_*=k/a(\eta_0)\simeq (-k\eta_0)H_{\rm inf}$,
because the physical momentum $k/a$ is larger than $M_\ast$
for $\eta<\eta_0$.

Before moving to an
explicit calculation of the bispectrum~(\ref{in-in}),
we comment on the crucial difference between
the calculation with the Bunch-Davies state and
that with non-Bunch-Davies initial states.
This difference explains the reason why we keep $\eta_0$ finite.
Formally, Eq.~(\ref{in-in}) includes an integral of the form:
\begin{align}
S(\tilde k):=\int^0_{\eta_0}\D\eta(-\eta)^ne^{ic_h\tilde k\eta}, \label{int-hhh}
\end{align}
where $n=1$ for the
standard cubic term with two spatial derivatives
and $n=2$ for the
$\dot h_{ij}\dot h_{jk}\dot h_{ki}$ term.

In the case of the Bunch-Davies initial state
in which there are only the positive-frequency modes
participating in this integral,
we have $\tilde k=k_t$ with
\begin{align}
k_t:=k_1+k_2+k_3>0,
\end{align}
and so
\begin{align}
S(\tilde k)\propto\frac{1}{(ic_h\tilde k)^{n+1}}, \label{int-BD}
\end{align}
because the exponential function rapidly oscillates
for $|c_h\tilde k\eta|\gg 1$.
In contrast to this standard case,
in the case of non-Bunch Davies states,
we have both positive and negative frequency modes
in the integral, leading to
$\tilde k=-k_m+k_{m+1}+k_{m+2}$ with $m$ being defined modulo $3$.
Note that
$\tilde k$ exactly vanishes at the flattened configuration,
$k_m=k_{m+1}+k_{m+2}$. For this configuration,
the exponential function no longer oscillates
even for $\eta\sim \eta_0$, and thus the integral reads
\begin{align}
S(\tilde k)\simeq\frac{(-\eta_0)^{n+1}}{n+1}, \label{FI}
\end{align}
which depends explicitly on $\eta_0$.
For the other configurations, the results of the integral are identical
to Eq.~(\ref{int-BD}).
In this section, we therefore need to calculate the primordial
bispectra in the two different cases separately,
the non-flattened and flattened configurations.

Let us define
\begin{align}
k_0:=k_1-k_2-k_3,
\end{align}
which appears frequently in the following discussion.

\subsection{Non-flattened configurations ($k_0\neq 0$)}

We first focus on the non-flattened configurations, i.e., $k_0\neq0$.
Assuming the de Sitter background and $M_T,\ c_h,\ M_4={\rm const}$,
the two contributions in the bispectrum~(\ref{3-pt}) respectively read
\begin{align}
\mathcal{B}^{s_1s_2s_3}_{({\rm GR})}&={\rm Re}\left[\tilde{\mathcal B}^{s_1s_2s_3}_{({\rm GR})}\right](s_1k_1+s_2k_2+s_3k_3)^2F(s_i,k_i), \label{3-ptgr}\\
\mathcal{B}^{s_1s_2s_3}_{({\rm new})}&={\rm Re}\left[\tilde{\mathcal{B}}^{s_1s_2s_3}_{(\rm new)}\right]F(s_i,k_i), \label{3-ptnew}
\end{align}
where
\begin{widetext}
\begin{align}
\tilde{\mathcal B}^{s_1s_2s_3}_{({\rm GR})}=&\frac{2H^4}{c_h^2M_T^4}\frac{1}{k_1^3k_2^3k_3^3}\left[\Pi_i\left(\alpha_{k_i}^{(s_i)*}-\beta_{k_i}^{(s_i)*}\right)\right]\biggl\{\left(\alpha_{k_1}^{(s_1)}\alpha_{k_2}^{(s_2)}\alpha_{k_3}^{(s_3)}+\beta_{k_1}^{(s_1)}\beta_{k_2}^{(s_2)}\beta_{k_3}^{(s_3)}\right)\mathcal{I}_0(k_1,k_2,k_3)\notag\\
&\ +\biggl[\biggl(\alpha_{k_1}^{(s_1)}\alpha_{k_2}^{(s_2)}\beta_{k_3}^{(s_3)}+\beta_{k_1}^{(s_1)}\beta_{k_2}^{(s_2)}\alpha_{k_3}^{(s_3)}\biggr)\mathcal{I}_1(k_1,k_2,k_3)+(k_1, s_1\leftrightarrow k_2,s_2)+(k_1, s_1\leftrightarrow k_3, s_3)\biggr]\biggr\}, \label{Bgr}\\
\tilde{\mathcal B}^{s_1s_2s_3}_{(\rm new)}=&\frac{192M_4{H^5}}{M_T^6}\frac{1}{k_1k_2k_3}\left[\Pi_i\left(\alpha_{k_i}^{(s_i)*}-\beta_{k_i}^{(s_i)*}\right)\right]\biggl\{\biggl(\alpha_{k_1}^{(s_1)}\alpha_{k_2}^{(s_2)}\alpha_{k_3}^{(s_3)}+\beta_{k_1}^{(s_1)}\beta_{k_2}^{(s_2)}\beta_{k_3}^{(s_3)}\biggr)\frac{1}{k_t^3}\notag\\
&\ -\biggl[\biggl(\alpha_{k_1}^{(s_1)}\beta_{k_2}^{(s_2)}\beta_{k_3}^{(s_3)}+\beta_{k_1}^{(s_1)}\alpha_{k_2}^{(s_2)}\alpha_{k_3}^{(s_3)}\biggr)\frac{1}{(-k_1+k_2+k_3)^3}+(k_1, s_1\leftrightarrow k_2, s_2)+(k_1, s_1\leftrightarrow k_3, s_3)\biggr]\biggr\}, \label{Bnew}
\end{align}
\end{widetext}
and
\begin{align}
F(s_i,k_i)&:=\frac{1}{64}\frac{k_t}{k_1^2k_2^2k_3^2}(s_1k_1+s_2k_2+s_3k_3)^2\notag\\
&\quad\ \times(k_1-k_2-k_3)(k_1+k_2-k_3)\notag\\
&\quad\ \times(k_1-k_2+k_3),\\
\mathcal{I}_0(k_1,k_2,k_3)&:=-k_t+\frac{k_1k_2k_3}{k_t^2}+\frac{k_1k_2+k_2k_3+k_3k_1}{k_t},\\
\mathcal{I}_1(k_1,k_2,k_3)&:=k_1+k_2-k_3+\frac{k_1k_2k_3}{(k_1+k_2-k_3)^2}\notag\\
&\quad\ +\frac{-k_1k_2+k_2k_3+k_1k_3}{k_1+k_2-k_3}.
\end{align}
These expressions are a generalization of Ref.~\cite{Gao:2011vs},
and reproduce the previous results by taking
the Bunch-Davies states ($\alpha^{(s)}_k=1$ and $\beta^{(s)}_k=0$).
Note that we have derived the auto-bispectrum from the $\dot h_{ij}\dot h_{jk}\dot h_{ki}$
term for the first time in the context of the non-Bunch-Davies states.

Taking into account the smallness of $\beta_k^{(s)}$
[Eqs.~(\ref{beta}) and~(\ref{alpha})], the resultant
bispectra to first order in $\beta^{(s)}_k$ are given by
\begin{align}
{\mathcal B}^{s_1s_2s_3}_{({\rm GR})}=&\frac{2H^4}{c_h^2M_T^4}\frac{1}{k_1^3k_2^3k_3^3}
(s_1k_1+s_2k_2+s_3k_3)^2F(s_i,k_i)\notag\\
&\times\biggl\{\bigl(1-\sum_i{\rm Re}[\beta_{k_i}^{(s_i)}]\bigr)\mathcal{I}_0(k_1,k_2,k_3)
\notag \\ & \quad \quad
+\left[{\rm Re}[\beta_{k_3}^{(s_3)}]\mathcal{I}_1(k_1,k_2,k_3)+\cdots\biggr]\right\},
 \label{Bgr-NBD}\\
{\mathcal B}^{s_1s_2s_3}_{(\rm new)}=&\frac{192M_4{H^5}}{M_T^6}
\frac{F(s_i,k_i)}{k_1k_2k_3}
\biggl\{\frac{1-\sum_i{\rm Re}[\beta_{k_i}^{(s_i)}]}{k_t^3}
\notag\\
&-\left[\frac{{\rm Re}[\beta_{k_1}^{(s_1)}]}{(-k_1+k_2+k_3)^3}+
\cdots\right]\biggr\}, \label{Bnew-NBD}
\end{align}
where the ellipsis denotes permutations.

Let us consider the squeezed configuration with
$k_L:=k_3\ll k_S:=k_1=k_2$. In the squeezed limit,
the expressions in the curly brackets in Eqs.~\eqref{Bgr-NBD} and~\eqref{Bnew-NBD}
are written respectively as
\begin{align}
\{\cdots\}&\simeq -\frac{3}{2}k_S\left(
1-\frac{4}{3}{\rm Re}[\beta_{k_S}^{(s_1)}+\beta_{k_S}^{(s_2)}]
\frac{k_S}{k_L}\right),\label{betasq1}
\end{align}
and
\begin{align}
\{\cdots\}&\simeq \frac{1}{8k_S^3}\left(
1-8 {\rm Re}[\beta_{k_S}^{(s_1)}+\beta_{k_S}^{(s_2)}]\frac{k_S^3}{k_L^3}\right).
\label{betasq2}
\end{align}
These equations show that
the effect of nonvanishing $\beta_k^{(s)}$ could be enhanced and seen in the
squeezed configuration.
In particular, the generation of squeezed non-Gaussianity
from the $\dot h_{ij}\dot h_{jk}\dot h_{ki}$ term
is in contrast with the standard case of
the Bunch-Davies state in which the bispectrum
has a peak at the equilateral
configuration~\cite{Gao:2011vs}.\footnote{Squeezed tensor non-Gaussianities
from the the $\dot h_{ij}\dot h_{jk}\dot h_{ki}$ operator
has been found also in the non-attractor inflation models~\cite{Ozsoy:2019slf}
and bouncing models~\cite{Akama:2019qeh}.}

To see whether this enhancement effect is significant or not,
let us take $k_S/k_L\sim 10^{2}$.
The non-Bunch-Davies contributions in Eqs.~(\ref{betasq1}) and~(\ref{betasq2})
are then of $\mathcal{O}(10^2|\beta^{(s)}_{k_S}|)$
and $\mathcal{O}(10^6|\beta^{(s)}_{k_S}|)$, respectively.
As arguned in Appendix~\ref{app},
the upper bound on the Bogoliubov coefficients
is obtained from the backreaction constraint,
which depends on the ratio $M_*/M_T(\sim M_*/\mpl)$.
If one takes $M_*\sim M_T\sim \mpl$, one has $|\beta_{k_S}^{(s)}|\lesssim 10^{-6}$,
so that the non-Bunch-Davies contribution in ${\cal B}_{{\rm (GR)}}^{s_1s_2s_3}$
is small, $\sim 10^{-4}$, while that in ${\cal B}_{{\rm (new)}}^{s_1s_2s_3}$
is of ${\cal O}(1)$.
This can be larger if one assumes smaller $M_*$.
For example, one gets $|\beta_{k_S}^{(s)}|\lesssim 10^{-2}$ if
$M_*\sim 10^{-2} M_T\sim 10^{-2} \mpl$.
In this case, the non-Bunch-Davies contribution
in ${\cal B}_{{\rm (GR)}}^{s_1s_2s_3}$ is of ${\cal O}(1)$
and that in ${\cal B}_{{\rm (new)}}^{s_1s_2s_3}$
is as large as ${\cal O}(10^{4})$.
Therefore, tensor squeezed non-Gaussianity could be
generated from the non-Bunch-Davies initial states,
depending on the parameters.

\subsection{Flattened Configuration ($k_0\to0$)}

So far we have assumed that $k_0=k_1-k_2-k_3\neq 0$.
Let us now investigate the flattened configuration, $k_0\simeq0$,
using Eq.~(\ref{FI}).
In this case, $\tilde{\mathcal B}^{s_1s_2s_3}_{({\rm GR})}$
and $\tilde{\mathcal B}^{s_1s_2s_3}_{({\rm new})}$
in Eqs.~(\ref{3-ptgr}) and~(\ref{3-ptnew}) are
given respectively by
\begin{align}
\tilde{\mathcal B}^{s_1s_2s_3}_{({\rm GR})}&\simeq\frac{2H^4}{c_h^2M_T^4}
\frac{1}{k_1^3k_2^3k_3^3}
\left[\Pi_i\left(\alpha_{k_i}^{(s_i)*}
-\beta_{k_i}^{(s_i)*}\right)\right]
\notag\\
&\quad\times\biggl[\biggl(\alpha_{k_1}^{(s_1)}
\alpha_{k_2}^{(s_2)}\alpha_{k_3}^{(s_3)}+\beta_{k_1}^{(s_1)}
\beta_{k_2}^{(s_2)}\beta_{k_3}^{(s_3)}\biggr)\notag\\
&\quad\quad\ \times\mathcal{I}_0(k_1,k_2,k_3)-\frac{k_1k_2k_3}{2}c_h^2\eta_0^2\notag\\
&\quad\quad\ \times\biggl(\beta^{(s_1)}_{k_1}
\alpha^{(s_2)}_{k_2}\alpha^{(s_3)}_{k_3}+\alpha^{(s_1)}_{k_1}
\beta^{(s_2)}_{k_2}\beta^{(s_3)}_{k_3}
\biggr)\biggr], \label{Bgr-f}\\
\tilde{\mathcal B}^{s_1s_2s_3}_{(\rm new)}&\simeq\frac{192M_4{H^5}}{M_T^6}
\frac{1}{k_1k_2k_3}\left[\Pi_i\left(\alpha_{k_i}^{(s_i)*}
-\beta_{k_i}^{(s_i)*}\right)\right]\notag\\
&\quad\times\biggl[\biggl(\alpha_{k_1}^{(s_1)}\alpha_{k_2}^{(s_2)}
\alpha_{k_3}^{(s_3)}+\beta_{k_1}^{(s_1)}\beta_{k_2}^{(s_2)}
\beta_{k_3}^{(s_3)}\biggr)\frac{1}{k_t^3}\notag\\
&\quad\quad\ +\frac{i}{6}c_h^3\eta_0^3
\biggl(\beta^{(s_1)}_{k_1}
\alpha^{(s_2)}_{k_2}\alpha^{(s_3)}_{k_3}-\alpha^{(s_1)}_{k_1}\beta^{(s_2)}_{k_2}\beta^{(s_3)}_{k_3}\biggr)
\biggr], \label{Bnew-f}
\end{align}
where we used $k_0\ll k_i,\ |c_hk_i\eta_0|\gg1$, and $|c_hk_0\eta_0|\ll1$.
In Ref.~\cite{Ashoorioon:2018sqb}, the flattened tensor non-Gaussianity
has already been studied,
but the interactions among the different polarization modes
have not been considered.

Similarly to the non-flattened configurations, we express the resultant
bispectra to first order in $\mathcal{O}(\beta^{(s)}_k)$ as
\begin{align}
  {\mathcal B}^{s_1s_2s_3}_{({\rm GR})}&\simeq\frac{2H^4}{c_h^2M_T^4}\frac{1}{k_1^3k_2^3k_3^3}(s_1k_1+s_2k_2+s_3k_3)^2
  F(s_i,k_i)\notag\\
  &\quad\times\biggl\{
\bigl(1-\sum_i{\rm Re}[\beta_{k_i}^{(s_i)}]\bigr)
  \mathcal{I}_0(k_1,k_2,k_3)
\notag \\ & \quad\quad
-\frac{k_1k_2k_3}{2}c_h^2\eta_0^2\,
{\rm Re}[\beta_{k_1}^{(s_1)}]\biggr\}, \label{Bgr-f2}\\
  {\mathcal B}^{s_1s_2s_3}_{(\rm new)}&
  \simeq\frac{192M_4{H^5}}{M_T^6}\frac{F(s_i,k_i)}{k_1k_2k_3}
  \notag \\ & \quad \times
  \biggl\{
  \frac{1-\sum_i{\rm Re}[\beta_{k_i}^{(s_i)}]}{k_t^3}
  -\frac{c_h^3\eta_0^3}{6}\,{\rm Im}[\beta_{k_1}^{(s_1)}]
  \biggr\}. \label{Bnew-f2}
\end{align}

Now we see that the primordial bispectra always vanish at the
exact flattened configurations,
because $F(s_i,k_i)=0$ for $k_0=0$.
This universal feature can be understood intuitively from
the viewpoint of angular momentum conservation~\cite{Agrawal:2018mrg}.
Although the expressions in the
curly brackets could be enhanced by powers of $k_i\eta_0$,
it would be difficult to obtain large flattened non-Gaussianities due
to this universal factor.\footnote{A different
conclusion was obtained in~\cite{Ashoorioon:2018sqb} because the
overall factor $F(s_i,k_i)$ was overlooked.}
This is in sharp contrast to the
result of the similar analysis for the curvature perturbation.
However, this is not the case for
the cross-interaction, as shown in the next section.

\section{Cross-bispectrum}

In this section, we consider a scalar-scalar-tensor bispectrum,
rather than a tensor-tensor-tensor bispectrum,
and explore the possibility of enhancing it with nontrivial initial states
of the tensor modes.
The cross-bispectrum we will consider is defined by
\begin{align}
&\langle0_b|\tilde\zeta(0,{\bf k}_1)\tilde\zeta(0,{\bf k}_2)\xi^{(s)}(0,{\bf k}_3)|0_b\rangle\notag\\
&=(2\pi)^3\delta({\bf k}_1+{\bf k}_2+{\bf k}_3)\mathcal{B}^{s}_{\zeta\zeta h}.
\end{align}

For the Lagrangian~\eqref{massL}, the quadratic
action for the curvature perturbation in the unitary gauge, $\zeta$,
takes the form~\cite{Kobayashi:2011nu}
\begin{align}
S_\zeta^{(2)}=\int\D t\D^3x\, \frac{a^3M_S^2}{c_s^2}
\left[\dot\zeta^2-\frac{c_s^2}{a^2}(\partial_i\zeta)^2\right],
\end{align}
where we do not need the concrete expression for $M_S$ and $c_s$
in the present discussion. These are time-dependent functions in general,
but in the inflationary universe we may assume that they are approximately constant.
We assume that
the Fourier component of the curvature perturbation, $\tilde \zeta(t,\mathbf{k})$,
can be written as
\begin{align}
\tilde\zeta = \psi_ka_\mathbf{k}+\psi^*_ka_{-\mathbf{k}}^\dagger,
\end{align}
where
\begin{align}
\psi_k&=
\frac{\sqrt{\pi}}{2\sqrt{2}a}\frac{c_s}{M_S}\sqrt{-\eta}H_{3/2}^{(1)}(-c_sk\eta)
\end{align}
is the Bunch-Davies mode function and the initial state is
in a vacuum state annihilated by $a_\mathbf{k}$.
By assuming this we focus on the effect of the excited tensor modes.

It has been found that the generic action
[Eqs.~(\ref{massS}) and~(\ref{massL})] introduces various cubic
operators that are not present in the simple case where
the inflaton is minimally coupled to gravity~\cite{Maldacena:2002vr}.
Among such operators
it is sufficient to consider one representative term that is
expected to be a
dominant source of the non-Gaussianities
in order to see whether the bispectrum can be enhanced or not.
Naively, operators with many derivatives are
important for the generation of non-Gaussianities
on subhorizon scales, and thus we focus on the
following interaction Hamiltonian:
\begin{align}
H^{\zeta\zeta h}_{\rm int}=-\int{}\D^3 x\frac{M_S^2\Lambda_{c}}{ac_s^2H^2}\partial^2 h_{ij}\partial_i\zeta\partial_j\zeta,
\end{align}
where we assume that $\Lambda_c={\rm const}$.
This term is indeed present in the
general Horndeski class of theories~\cite{Gao:2012ib}.

Similarly to the auto-correlation function, the cross-correlation function includes
the integral
\begin{align}
S_c(\tilde k_c):=\int^0_{\eta_0}\D\eta(-\eta)^3e^{i\tilde k_c\eta}, \label{int-ssh}
\end{align}
where
\begin{align}
\tilde k_c:=c_h k_3-c_s(k_1+k_2). \label{ks}
\end{align}
For the configuration satisfying $\tilde k_c=0$,
the cross-bispectrum depends on $\eta_0$ and
is enhanced by powers of $k_i\eta_0$ due to the excited
tensor modes. Note that
this configuration depends on the propagation speeds.  
For given $c_s/c_h\,(<1)$, one has a one-parameter family of
different shapes satisfying $\tilde k_c=0$ away from the flattened configuration.

In the same way as the previous calculations,
we derive the cross-bispectrum to first order in $\beta^{(s)}_{k_3}$:
\begin{align}
\mathcal{B}^s_{\zeta\zeta h}&=\left.\mathcal{B}^s_{\zeta\zeta h,({\rm BD})}
\right|_{\tilde k_c=0}\notag\\
&\quad\times\biggl
\{1-{\rm Re}[\beta^{(s)}_{k_3}]\notag\\
&\quad\quad\ +\frac{(2/5)k_1k_2}{2k_1^2+5k_1k_2+2k_2^2}c_s^4(k_1+k_2)^4\eta_0^4
{\rm Re}[\beta^{(s)}_{k_3}]\biggr\},
\end{align}
where $\mathcal{B}^{s}_{\zeta\zeta h}$
is the cross-bispectrum in the case of the Bunch-Davies initial state.
This quantity is obtained in~\cite{Gao:2012ib} as
\begin{align}
\mathcal{B}^s_{\zeta\zeta h,({\rm BD})}&=
\frac{H^4\Lambda_c}{M_S^2M_T^2c_s^4c_h}\cdot
\frac{k_t}{16k_1^3k_2^3k_3^3}
\notag \\
&\quad \times \frac{(k_1-k_2-k_3)(k_1+k_2-k_3)(k_1-k_2+k_3)}{[c_s(k_1+k_2)+c_hk_3]^4}
\notag \\
& \quad \times
\biggl\{
c_s^2[c_s(k_1+k_2)+4c_hk_3](k_1^2+3k_1k_2+k_2^2)
\notag\\
&\quad \quad+c_h^2k_3^2[4c_s(k_1+k_2)+c_hk_3]\biggr\}.
\end{align}
From the above result we see
that the non-Bunch-Davies contribution
is of $\mathcal{O}(\beta^{(s)}_kc_s^4k_i^4\eta_0^4)$.

In the actual observables, we anticipate that
this non-Bunch-Davies enhancement will be softened
by (at least) one power of $|k\eta_0|$ due to the angular averaging~\cite{Holman:2007na}.
Let us therefore estimate roughly how large $\beta^{(s)}_k(c_sk_i\eta_0)^n$
could be. As argued in Appendix~\ref{app}, the Bogoliubov coefficients
have an upper bound from the backreaction constraint,
which depends on the cutoff scale. We also have
$|c_sk_i\eta_0| \lesssim c_s M_*/H_{\rm inf}$.
Combining these, we find
\begin{align}
\beta^{(s)}_k(c_sk_i\eta_0)^n\lesssim
\frac{c_s^n}{c_h^{1/2}}\frac{\mpl M_*^{n-2}}{H_{\rm inf}^{n-1}}.
\end{align}
Even for $n=2$ the upper bound is typically larger than ${\cal O}(1)$.
We thus conclude that initially excited tensor modes can leave
a potentially observable imprint in the cross-bispectrum\footnote{In the present paper, we have considered the scalar-scalar-tensor bispectrum, but initially excited scalar modes would be able to enhance the scalar-tensor-tensor bispectrum as well.}.

\section{Summary}

In the present paper, we have considered primordial tensor perturbations
with non-Bunch-Davies initial states.
Employing a general scalar-tensor theory,
we have described non-minimal couplings between gravity and the inflaton.

First, we evaluated the size of tensor three-point functions
and showed that the squeezed non-Gaussianities
in particular from the newly introduced operator in
non-minimally coupled theories
can potentially be enhanced.
In contrast to the case of the scalar three-point functions~\cite{Holman:2007na},
the tensor three-point function always vanishes at the flattened
momentum triangles. This is as it should be,
as can be seen from the momentum conservation argument~\cite{Agrawal:2018mrg}.

Next, we have studied the cross-bispectrum involving
one tensor and two scalar modes.
We have found that the enhancement due to
the non-Bunch-Davies effect can be large at non-trivial
triangle shapes. It whould therefore be interesting
to investigate how such non-Gaussian signature is imprinted
e.g. on CMB bispectra~\cite{Tahara:2017wud}, which we leave for further studies.

\acknowledgments
We would like to thank Emanuela Dimastrogiovanni and Tomohiro Fujita for helpful comments. 
The work of SA was supported by the JSPS Research Fellowships for Young Scientists
No.~18J22305. The work of SH was supported by
the JSPS Research Fellowships for Young Scientists No.~17J04865.
The work of TK was supported by
MEXT KAKENHI Grant Nos.~JP17H06359, JP16K17707, and JP18H04355.


\appendix

\section{Backreaction constraint on $\beta^{(s)}_k$}\label{app}

If a scalar field is minimally coupled to gravity,
the energy-momentum tensor of tensor perturbations
is derived by expanding the Einstein tensor
to second order in $h_{ij}$.
Even if the scalar field is non-minimally coupled to gravity,
one may proceed essentially in the same way and expand the field equations
to second order in $h_{ij}$ to estimate the energy density of tensor
perturbations. This is how one can evaluate the backreaction of
excited tensor modes to the homogeneous background.
The effective energy density of
subhorizon tensor perturbations is thus given by
\begin{align}
\rho_h\sim \frac{M_T^2}{a^2c_h^2}{h'_{ij}}^2
\sim M_T^2\frac{(\partial_ih_{jk})^2}{a^2},
\end{align}
where a dash stands for differentiation with respect to $\eta$.
The backreaction can safely be ignored if
\begin{align}
\langle 0_b|\hat \rho_h|0_b\rangle \lesssim \bar{\cal E},
\end{align}
where $\bar{\cal E}$ is the homogeneous part of the
field equation, which can be estimated naively as
\begin{align}
\bar{\cal E}\sim \mpl^2H_{\rm inf}^2,
\end{align}
where $H_{\rm inf}$ is the inflationary Hubble parameter and
$\mpl\sim M_T$.

The backreaction from the excited modes of tensor perturbations
can be estimated at $\eta=\eta_0$ from
\begin{align}
\langle0_b|\hat\rho_h|0_b\rangle&\sim
\frac{M_T^2}{a^2c_h^2}\langle0_b|{\hat h_{ij}'^2}|0_b\rangle
\notag \\ &\sim
\frac{c_h}{a^4(\eta_0)}\int^{M_*a(\eta_0)}_0 |\beta^{(s)}_k|^2k^3\D k
,\label{br}
\end{align}
where we discarded the vacuum energy.
Then, by requiring that
\begin{align}
\frac{c_h}{a^4(\eta_0)}\int^{M_*a(\eta_0)}_0 |\beta^{(s)}_k|^2k^3\D k
 \lesssim \mpl^2H_{\rm inf}^2, \label{back}
\end{align}
one can save the inflationary background from being spoiled by the backreaction.

To derive a more explicit constraint,
we need to assume the momentum dependence
of the Bogoliubov coefficients.
Here, let us suppose that $\beta^{(s)}_k$ is of the form
\begin{align}
\beta^{(s)}_k\sim \beta \exp\left[-\frac{k^2}{M_*^2a^2(\eta_0)}\right]
\end{align}
as a simple model,
where $\beta$ is a constant parameter.
Substituting this into Eq.~(\ref{back}), we obtain
\begin{align}
|\beta|^2
\lesssim\frac{1}{c_h}\left(\frac{\mpl}{M_*}\right)^2
\left(\frac{H_{\rm inf}}{M_*}\right)^2. \label{constb}
\end{align}
As is explained in the main text,
the deviation of the tensor power spectrum from the standard
Bunch-Davies result
is at most of
$\mathcal{O}(|\beta^{(s)}_k|)\ll1$,
and thus we may use $\mathcal{P}_h\sim H_{\rm inf}^2/(c_hM_T^2)$.
Then,
the constraint~(\ref{constb}) can be rewritten as
\begin{align}
|\beta|^2&
\lesssim\mathcal{P}_h\frac{\mpl^2}{M_*^2}
\frac{M_T^2}{M_*^2}\sim r\mathcal{P}_\zeta
\frac{\mpl^2M_T^2}{M_*^4} \notag\\
&\lesssim 10^{-11} \frac{\mpl^2M_T^2}{M_*^4} .
\end{align}
For example, if we take $M_*\sim \mpl\sim M_T$,
then we have $|\beta|\lesssim 10^{-6}$, while
 if we assume that the cutoff scale is much smaller, say,
$M_*\sim 10^{-2}\mpl\sim 10^{-2}M_T$,
the bound is looser, $|\beta|\lesssim 10^{-2}$.


\begin{thebibliography}{99}

\bibitem{Guth:1980zm}
  A.~H.~Guth,
  ``The Inflationary Universe: A Possible Solution to the Horizon and Flatness Problems,''
  Phys.\ Rev.\ D {\bf 23}, 347 (1981).

\bibitem{Starobinsky:1980te}
  A.~A.~Starobinsky,
  ``A New Type of Isotropic Cosmological Models Without Singularity,''
  Phys.\ Lett.\ B {\bf 91}, 99 (1980).

\bibitem{Sato:1980yn}
  K.~Sato,
  ``First Order Phase Transition of a Vacuum and Expansion of the Universe,''
  Mon.\ Not.\ Roy.\ Astron.\ Soc.\  {\bf 195}, 467 (1981).


\bibitem{Akrami:2018odb}
  Y.~Akrami {\it et al.} [Planck Collaboration],
  ``Planck 2018 results. X. Constraints on inflation,''
  arXiv:1807.06211 [astro-ph.CO].


\bibitem{Bunch:1978yq} 
  T.~S.~Bunch and P.~C.~W.~Davies,
  ``Quantum Field Theory in de Sitter Space: Renormalization by Point Splitting,''
  Proc.\ Roy.\ Soc.\ Lond.\ A {\bf 360}, 117 (1978).



%













\bibitem{Chen:2006nt}
  X.~Chen, M.~x.~Huang, S.~Kachru and G.~Shiu,
  ``Observational signatures and non-Gaussianities of general single field inflation,''
  JCAP {\bf 0701}, 002 (2007)
  [hep-th/0605045].

\bibitem{Holman:2007na}
  R.~Holman and A.~J.~Tolley,
  ``Enhanced Non-Gaussianity from Excited Initial States,''
  JCAP {\bf 0805}, 001 (2008)
  [arXiv:0710.1302 [hep-th]].
  

\bibitem{Xue:2008mk}
  W.~Xue and B.~Chen,
  ``alpha-vacuum and inflationary bispectrum,''
  Phys.\ Rev.\ D {\bf 79}, 043518 (2009)
  [arXiv:0806.4109 [hep-th]].

\bibitem{Meerburg:2009ys}
  P.~D.~Meerburg, J.~P.~van der Schaar and P.~S.~Corasaniti,
  ``Signatures of Initial State Modifications on Bispectrum Statistics,''
  JCAP {\bf 0905}, 018 (2009)
  [arXiv:0901.4044 [hep-th]].

\bibitem{Meerburg:2009fi}
  P.~D.~Meerburg, J.~P.~van der Schaar and M.~G.~Jackson,
  ``Bispectrum signatures of a modified vacuum in single field inflation with a small speed of sound,''
  JCAP {\bf 1002}, 001 (2010)
  [arXiv:0910.4986 [hep-th]].

\bibitem{Chen:2010xka} 
  X.~Chen,
  ``Primordial Non-Gaussianities from Inflation Models,''
  Adv.\ Astron.\  {\bf 2010}, 638979 (2010)
  [arXiv:1002.1416 [astro-ph.CO]].
  
\bibitem{Meerburg:2010ks} 
  P.~D.~Meerburg,
  ``Oscillations in the Primordial Bispectrum I: Mode Expansion,''
  Phys.\ Rev.\ D {\bf 82}, 063517 (2010)
  [arXiv:1006.2771 [astro-ph.CO]].
  
  Chen:2010bka,Meerburg:2010rp,
      
\bibitem{Chen:2010bka}
  X.~Chen,
  ``Folded Resonant Non-Gaussianity in General Single Field Inflation,''
  JCAP {\bf 1012}, 003 (2010)
  [arXiv:1008.2485 [hep-th]].

\bibitem{Meerburg:2010rp}
  P.~D.~Meerburg and J.~P.~van der Schaar,
  ``Minimal cut-off vacuum state constraints from CMB bispectrum statistics,''
  Phys.\ Rev.\ D {\bf 83}, 043520 (2011)
  [arXiv:1009.5660 [hep-th]].

\bibitem{Agullo:2010ws} 
  I.~Agullo and L.~Parker,
  ``Non-gaussianities and the Stimulated creation of quanta in the inflationary universe,''
  Phys.\ Rev.\ D {\bf 83}, 063526 (2011)
  [arXiv:1010.5766 [astro-ph.CO]].
  

    
\bibitem{Ashoorioon:2010xg}
  A.~Ashoorioon and G.~Shiu,
  ``A Note on Calm Excited States of Inflation,''
  JCAP {\bf 1103}, 025 (2011)
  [arXiv:1012.3392 [astro-ph.CO]].

\bibitem{Ganc:2011dy}
  J.~Ganc,
  ``Calculating the local-type fNL for slow-roll inflation with a non-vacuum initial state,''
  Phys.\ Rev.\ D {\bf 84}, 063514 (2011)
  [arXiv:1104.0244 [astro-ph.CO]].

\bibitem{Ganc:2012ae} 
  J.~Ganc and E.~Komatsu,
  ``Scale-dependent bias of galaxies and mu-type distortion of the cosmic microwave background spectrum from single-field inflation with a modified initial state,''
  Phys.\ Rev.\ D {\bf 86}, 023518 (2012)
  [arXiv:1204.4241 [astro-ph.CO]].
  
\bibitem{Agarwal:2012mq}
  N.~Agarwal, R.~Holman, A.~J.~Tolley and J.~Lin,
  ``Effective field theory and non-Gaussianity from general inflationary states,''
  JHEP {\bf 1305}, 085 (2013)
  [arXiv:1212.1172 [hep-th]].

\bibitem{Gong:2013yvl}
  J.~O.~Gong and M.~Sasaki,
  ``Squeezed primordial bispectrum from general vacuum state,''
  Class.\ Quant.\ Grav.\  {\bf 30}, 095005 (2013)
  [arXiv:1302.1271 [astro-ph.CO]].

\bibitem{Flauger:2013hra} 
  R.~Flauger, D.~Green and R.~A.~Porto,
  ``On squeezed limits in single-field inflation.  Part I,''
  JCAP {\bf 1308}, 032 (2013)
  [arXiv:1303.1430 [hep-th]].
  
    
\bibitem{Aravind:2013lra}
  A.~Aravind, D.~Lorshbough and S.~Paban,
  ``Non-Gaussianity from Excited Initial Inflationary States,''
  JHEP {\bf 1307}, 076 (2013)
  [arXiv:1303.1440 [hep-th]].

\bibitem{Ashoorioon:2013eia} 
  A.~Ashoorioon, K.~Dimopoulos, M.~M.~Sheikh-Jabbari and G.~Shiu,
  ``Reconciliation of High Energy Scale Models of Inflation with Planck,''
  JCAP {\bf 1402}, 025 (2014)
  [arXiv:1306.4914 [hep-th]].


\bibitem{Brahma:2013rua}
  S.~Brahma, E.~Nelson and S.~Shandera,
  ``Fossilized Gravitational Wave Relic and Primordial Clocks,''
  Phys.\ Rev.\ D {\bf 89}, no. 2, 023507 (2014)
  [arXiv:1310.0471 [astro-ph.CO]].

\bibitem{Bahrami:2013isa}
  S.~Bahrami and E.~E.~Flanagan,
  ``Primordial non-Gaussianities in single field inflationary models with non-trivial initial states,''
  JCAP {\bf 1410}, 010 (2014)
  [arXiv:1310.4482 [astro-ph.CO]].

\bibitem{Kundu:2013gha}
  S.~Kundu,
  ``Non-Gaussianity Consistency Relations, Initial States and Back-reaction,''
  JCAP {\bf 1404}, 016 (2014)
  [arXiv:1311.1575 [astro-ph.CO]].
  
\bibitem{Emami:2014tpa} 
  R.~Emami, H.~Firouzjahi and M.~Zarei,
  ``Anisotropic inflation with the nonvacuum initial state,''
  Phys.\ Rev.\ D {\bf 90}, no. 2, 023504 (2014)
  [arXiv:1401.4406 [hep-th]].

\bibitem{Zeynizadeh:2015eia} 
  S.~Zeynizadeh and A.~R.~Akbarieh,
  ``Higgs Inflation and General Initial Conditions,''
  Eur.\ Phys.\ J.\ C {\bf 75}, no. 8, 355 (2015)
  [arXiv:1504.00482 [hep-th]].
  
\bibitem{Meerburg:2015yka} 
  P.~D.~Meerburg and M.~M{$\ddot u$}nchmeyer,
  ``Optimal CMB estimators for bispectra from excited states,''
  Phys.\ Rev.\ D {\bf 92}, no. 6, 063527 (2015)
  [arXiv:1505.05882 [astro-ph.CO]].
    
\bibitem{Ashoorioon:2016lrg} 
  A.~Ashoorioon, R.~Casadio and T.~Koivisto,
  ``Anisotropic non-Gaussianity from Rotational Symmetry Breaking Excited Initial States,''
  JCAP {\bf 1612}, 002 (2016)
  [arXiv:1605.04758 [hep-th]].  
    
\bibitem{Shukla:2016bnu} 
  A.~Shukla, S.~P.~Trivedi and V.~Vishal,
  ``Symmetry constraints in inflation, $\alpha$-vacua, and the three point function,''
  JHEP {\bf 1612}, 102 (2016)
  [arXiv:1607.08636 [hep-th]].  
  
\bibitem{Ashoorioon:2018sqb}
  A.~Ashoorioon,
  ``Rescuing Single Field Inflation from the Swampland,''
  Phys.\ Lett.\ B {\bf 790}, 568 (2019)
  [arXiv:1810.04001 [hep-th]].
  




\bibitem{Gleyzes:2014dya} 
  J.~Gleyzes, D.~Langlois, F.~Piazza and F.~Vernizzi,
  ``Healthy theories beyond Horndeski,''
  Phys.\ Rev.\ Lett.\  {\bf 114}, no. 21, 211101 (2015)
  [arXiv:1404.6495 [hep-th]].
  
\bibitem{Gao:2014soa} 
  X.~Gao,
  ``Unifying framework for scalar-tensor theories of gravity,''
  Phys.\ Rev.\ D {\bf 90}, 081501 (2014)
  [arXiv:1406.0822 [gr-qc]].
  
  
  
    
\bibitem{Kobayashi:2011nu}
  T.~Kobayashi, M.~Yamaguchi and J.~Yokoyama,
  ``Generalized G-inflation: Inflation with the most general second-order field equations,''
  Prog.\ Theor.\ Phys.\  {\bf 126}, 511 (2011)
  [arXiv:1105.5723 [hep-th]].
  
\bibitem{Gao:2011vs}
  X.~Gao, T.~Kobayashi, M.~Yamaguchi and J.~Yokoyama,
  ``Primordial non-Gaussianities of gravitational waves in the most general single-field inflation model,''
  Phys.\ Rev.\ Lett.\  {\bf 107}, 211301 (2011)
  [arXiv:1108.3513 [astro-ph.CO]].  
  
\bibitem{Ozsoy:2019slf} 
  O.~Ozsoy, M.~Mylova, S.~Parameswaran, C.~Powell, G.~Tasinato and I.~Zavala,
  ``Squeezed tensor non-Gaussianity in non-attractor inflation,''
  JCAP {\bf 1909}, no. 09, 036 (2019)
  [arXiv:1902.04976 [hep-th]].
  
\bibitem{Akama:2019qeh} 
  S.~Akama, S.~Hirano and T.~Kobayashi,
  ``Primordial non-Gaussianities of scalar and tensor perturbations in general bounce cosmology: Evading the no-go theorem,''
  Phys.\ Rev.\ D {\bf 101}, no. 4, 043529 (2020)
  [arXiv:1908.10663 [gr-qc]].

\bibitem{Giovannini:2015kfa} 
  M.~Giovannini,
  ``The refractive index of relic gravitons,''
  Class.\ Quant.\ Grav.\  {\bf 33}, no. 12, 125002 (2016)
  [arXiv:1507.03456 [astro-ph.CO]].

  \bibitem{Cai:2015yza}
    Y.~Cai, Y.~T.~Wang and Y.~S.~Piao,
    ``Is there an effect of a nontrivial $c_T$ during inflation?,''
    Phys.\ Rev.\ D {\bf 93}, no. 6, 063005 (2016)
    [arXiv:1510.08716 [astro-ph.CO]].

    \bibitem{Cai:2016ldn}
  Y.~Cai, Y.~T.~Wang and Y.~S.~Piao,
  ``Propagating speed of primordial gravitational waves and inflation,''
  Phys.\ Rev.\ D {\bf 94}, no. 4, 043002 (2016)
  [arXiv:1602.05431 [astro-ph.CO]].
  
\bibitem{Giovannini:2018zbf} 
  M.~Giovannini,
  ``The propagating speed of relic gravitational waves and their refractive index during inflation,''
  Eur.\ Phys.\ J.\ C {\bf 78}, no. 6, 442 (2018)
  [arXiv:1803.05203 [gr-qc]].

\bibitem{Giovannini:2018nkt} 
  M.~Giovannini,
  ``Blue and violet graviton spectra from a dynamical refractive index,''
  Phys.\ Lett.\ B {\bf 789}, 502 (2019)
  [arXiv:1805.08142 [astro-ph.CO]].
  
  

  \bibitem{Ashoorioon:2014nta}
    A.~Ashoorioon, K.~Dimopoulos, M.~M.~Sheikh-Jabbari and G.~Shiu,
    ``Non-Bunch-Davis initial state reconciles chaotic models with BICEP and Planck,''
    Phys.\ Lett.\ B {\bf 737}, 98 (2014)
    [arXiv:1403.6099 [hep-th]].
    
\bibitem{Agrawal:2018mrg}
  A.~Agrawal, T.~Fujita and E.~Komatsu,
  ``Tensor Non-Gaussianity from Axion-Gauge-Fields Dynamics : Parameter Search,''
  JCAP {\bf 1806}, 027 (2018)
  [arXiv:1802.09284 [astro-ph.CO]].  
    
\bibitem{Maldacena:2002vr} 
  J.~M.~Maldacena,
  ``Non-Gaussian features of primordial fluctuations in single field inflationary models,''
  JHEP {\bf 0305}, 013 (2003)
  [astro-ph/0210603].
      

\bibitem{Gao:2012ib}
  X.~Gao, T.~Kobayashi, M.~Shiraishi, M.~Yamaguchi, J.~Yokoyama and S.~Yokoyama,
  ``Full bispectra from primordial scalar and tensor perturbations in the most general single-field inflation model,''
  PTEP {\bf 2013}, 053E03 (2013)
  [arXiv:1207.0588 [astro-ph.CO]].

\bibitem{Tahara:2017wud}
  H.~W.~H.~Tahara and J.~Yokoyama,
  ``CMB B-mode auto-bispectrum produced by primordial gravitational waves,''
  PTEP {\bf 2018}, no. 1, 013E03 (2018)
  [arXiv:1704.08904 [astro-ph.CO]].

\end{thebibliography}
\end{document}